\pdfoutput=1

\documentclass[12pt,a4paper]{article}

\usepackage{lineno}  
\usepackage{graphicx}  

\usepackage{xspace} 
\usepackage{color}
\usepackage{colortbl}
\usepackage{amsmath} 
\usepackage{ifthen} 
\newboolean{pdflatex}
\setboolean{pdflatex}{true} 

\newboolean{articletitles}
\setboolean{articletitles}{true} 

\newboolean{uprightparticles}
\setboolean{uprightparticles}{false} 

\usepackage{amssymb}
\usepackage{amsfonts}
\usepackage{upgreek} 

\usepackage{hyperref}    
\usepackage[all]{hypcap} 

\def\lhcb {LHCb\xspace}
\def\ux85 {UX85\xspace}

\ifthenelse{\boolean{uprightparticles}}%
{

 \def\Ppi         {\ensuremath{\uppi}\xspace}

 \def\PDelta      {\ensuremath{\Delta}\xspace}                 
 \def\PXi      {\ensuremath{\Xi}\xspace}                 
 \def\PLambda      {\ensuremath{\Lambda}\xspace}                 
 \def\PSigma      {\ensuremath{\Sigma}\xspace}                 
 \def\POmega      {\ensuremath{\Omega}\xspace}                 
 \def\PUpsilon      {\ensuremath{\Upsilon}\xspace}                 
 
 \def\PB      {\ensuremath{\mathrm{B}}\xspace}                 
                  
 \def\PD      {\ensuremath{\mathrm{D}}\xspace}

 \def\PK      {\ensuremath{\mathrm{K}}\xspace}

 \def\Pb      {\ensuremath{\mathrm{b}}\xspace}                 
 \def\Pc      {\ensuremath{\mathrm{c}}\xspace}

 \def\Pi      {\ensuremath{\mathrm{i}}\xspace}

 \def\Ps      {\ensuremath{\mathrm{s}}\xspace}

}
{

 \def\Ppi         {\ensuremath{\pi}\xspace}

 \mathchardef\PDelta="7101
 \mathchardef\PXi="7104
 \mathchardef\PLambda="7103
 \mathchardef\PSigma="7106
 \mathchardef\POmega="710A
 \mathchardef\PUpsilon="7107
                  
 \def\PB      {\ensuremath{B}\xspace}                 
                  
 \def\PD      {\ensuremath{D}\xspace}

 \def\PK      {\ensuremath{K}\xspace}

 \def\Pb      {\ensuremath{b}\xspace}                 
 \def\Pc      {\ensuremath{c}\xspace}

 \def\Pi      {\ensuremath{i}\xspace}

 \def\Ps      {\ensuremath{s}\xspace}

}

\def\squark    {\ensuremath{\Ps}\xspace}

\def\cquark    {\ensuremath{\Pc}\xspace}

\def\bquark    {\ensuremath{\Pb}\xspace}

\def\pion  {\ensuremath{\Ppi}\xspace}

\def\pip   {\ensuremath{\pion^+}\xspace}
\def\pim   {\ensuremath{\pion^-}\xspace}
\def\pipi  {\ensuremath{\pion^+\pion^-}\xspace}
\def\pipm  {\ensuremath{\pion^\pm}\xspace}

\def\kaon  {\ensuremath{\PK}\xspace}
  \def\Kbar  {\kern 0.2em\overline{\kern -0.2em \PK}{}\xspace}

\def\Kz    {\ensuremath{\kaon^0}\xspace}
\def\Kzb   {\ensuremath{\Kbar^0}\xspace}
\def\KzKzb {\ensuremath{\Kz \kern -0.16em \Kzb}\xspace}
\def\Kp    {\ensuremath{\kaon^+}\xspace}
\def\Km    {\ensuremath{\kaon^-}\xspace}

\def\KpKm  {\ensuremath{\Kp \kern -0.16em \Km}\xspace}

  \def\Dbar    {\kern 0.2em\overline{\kern -0.2em \PD}{}\xspace}
\def\D       {\ensuremath{\PD}\xspace}

\def\Dz      {\ensuremath{\D^0}\xspace}
\def\Dzb     {\ensuremath{\Dbar^0}\xspace}
\def\DzDzb   {\ensuremath{\Dz {\kern -0.16em \Dzb}}\xspace}
\def\Dp      {\ensuremath{\D^+}\xspace}
\def\Dm      {\ensuremath{\D^-}\xspace}

\def\DpDm    {\ensuremath{\Dp {\kern -0.16em \Dm}}\xspace}

\def\Dstarp  {\ensuremath{\D^{*+}}\xspace}
\def\Dstarm  {\ensuremath{\D^{*-}}\xspace}
\def\Dstarpm {\ensuremath{\D^{*\pm}}\xspace}

\def\Ds      {\ensuremath{\D^+_\squark}\xspace}
\def\Dsp     {\ensuremath{\D^+_\squark}\xspace}
\def\Dsm     {\ensuremath{\D^-_\squark}\xspace}
\def\Dspm    {\ensuremath{\D^{\pm}_\squark}\xspace}

  \def\Bbar    {\kern 0.18em\overline{\kern -0.18em \PB}{}\xspace}

  \def\Y#1S{\ensuremath{\PUpsilon{(#1S)}}\xspace}

\def\to                 {\ensuremath{\rightarrow}\xspace}

\def\CP                {\ensuremath{C\!P}\xspace}

\def\AT#1     {\ensuremath{A_{\mathrm{T}}^{#1}}\xspace}           

\def\C#1      {\ensuremath{\mathcal{C}_{#1}}\xspace}                       
\def\Cp#1     {\ensuremath{\mathcal{C}_{#1}^{'}}\xspace}                    
\def\Ceff#1   {\ensuremath{\mathcal{C}_{#1}^{\mathrm{(eff)}}}\xspace}        
\def\Cpeff#1  {\ensuremath{\mathcal{C}_{#1}^{'\mathrm{(eff)}}}\xspace}       
\def\Ope#1    {\ensuremath{\mathcal{O}_{#1}}\xspace}                       
\def\Opep#1   {\ensuremath{\mathcal{O}_{#1}^{'}}\xspace}                    

\newcommand{\tev}{\ensuremath{\mathrm{\,Te\kern -0.1em V}}\xspace}
\newcommand{\gev}{\ensuremath{\mathrm{\,Ge\kern -0.1em V}}\xspace}
\newcommand{\mev}{\ensuremath{\mathrm{\,Me\kern -0.1em V}}\xspace}
\newcommand{\kev}{\ensuremath{\mathrm{\,ke\kern -0.1em V}}\xspace}
\newcommand{\ev}{\ensuremath{\mathrm{\,e\kern -0.1em V}}\xspace}
\newcommand{\gevc}{\ensuremath{{\mathrm{\,Ge\kern -0.1em V\!/}c}}\xspace}
\newcommand{\mevc}{\ensuremath{{\mathrm{\,Me\kern -0.1em V\!/}c}}\xspace}
\newcommand{\gevcc}{\ensuremath{{\mathrm{\,Ge\kern -0.1em V\!/}c^2}}\xspace}
\newcommand{\gevgevcccc}{\ensuremath{{\mathrm{\,Ge\kern -0.1em V^2\!/}c^4}}\xspace}
\newcommand{\mevcc}{\ensuremath{{\mathrm{\,Me\kern -0.1em V\!/}c^2}}\xspace}

\def\mub{\ensuremath{\rm \,\upmu b}\xspace}

\newcommand{\chisq}{\ensuremath{\chi^2}\xspace}

\def\gsim{{~\raise.15em\hbox{$>$}\kern-.85em
          \lower.35em\hbox{$\sim$}~}\xspace}
\def\lsim{{~\raise.15em\hbox{$<$}\kern-.85em
          \lower.35em\hbox{$\sim$}~}\xspace}

\def\ptot       {\mbox{$p$}\xspace}
\def\pt         {\mbox{$p_{\rm T}$}\xspace}

\def\tell1  {TELL1\xspace}
\def\ukl1   {UKL1\xspace}

\usepackage{mciteplus}

\begin{document}
\renewcommand{\thefootnote}{\fnsymbol{footnote}}
\setcounter{footnote}{1}


\begin{titlepage}
\pagenumbering{roman}
\belowpdfbookmark{Title page}{title}

\pagenumbering{roman}
\vspace*{-1.5cm}
\centerline{\large EUROPEAN ORGANIZATION FOR NUCLEAR RESEARCH (CERN)}
\vspace*{1.5cm}
\hspace*{-5mm}\begin{tabular*}{16cm}{lc@{\extracolsep{\fill}}r}
\vspace*{-12mm}\mbox{\!\!\!\includegraphics[width=.12\textwidth]{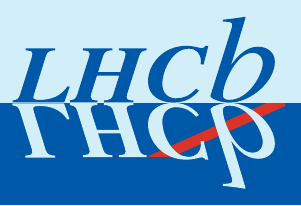}}& & \\
 & & CERN-PH-EP-2012-114\\
 & & LHCb-PAPER-2012-009 \\  
 & & May 4, 2012 \\ 
 & & \\
\hline
\end{tabular*}

\vspace*{2.0cm}

{\bf\boldmath\huge
\begin{center}
Measurement of the $\Dsp-\Dsm$ production asymmetry in 7~TeV $pp$ collisions
\end{center}
}

\vspace*{1.0cm}
\begin{center}
\normalsize {The LHCb collaboration\footnote{Authors are listed on the following pages.}
}
\end{center}

\vspace{\fill}

\begin{abstract}

Heavy quark production in 7~TeV centre-of-mass energy $pp$ collisions at the LHC is not necessarily flavour symmetric. The production asymmetry, $A_{\rm P}$, between $D_s^+$ and $D_s^-$ mesons is studied using the $\phi\pi^{\pm}$ decay mode in a data sample of 1.0~fb$^{-1}$ collected with the LHCb detector. 
The difference between $\pi^+$ and $\pi^-$ detection efficiencies is determined using the ratios of fully reconstructed to partially reconstructed $\Dstarpm$ decays. 
The overall production asymmetry in the $D_s^{\pm}$ rapidity region 2.0 to 4.5 with transverse momentum larger than 2\,GeV
is measured to be $A_{\rm P}=(-0.33\pm0.22 \pm0.10)$\%. This result can constrain models of heavy flavour production.

\end{abstract}

\vspace*{1.5cm}

\vspace*{1.0cm}

\vspace*{2.0cm}

\begin{center}
Submitted to Physics Letters B
\end{center}

\vspace{\fill}

\newpage
\centerline{\large\bf LHCb collaboration}
\begin{flushleft}
\small
R.~Aaij$^{38}$, 
C.~Abellan~Beteta$^{33,n}$, 
A.~Adametz$^{11}$, 
B.~Adeva$^{34}$, 
M.~Adinolfi$^{43}$, 
C.~Adrover$^{6}$, 
A.~Affolder$^{49}$, 
Z.~Ajaltouni$^{5}$, 
J.~Albrecht$^{35}$, 
F.~Alessio$^{35}$
% ...

{\footnotesize \it
$ ^{1}$Centro Brasileiro de Pesquisas F\'{i}sicas (CBPF), Rio de Janeiro, Brazil\\
% ...
$ ^{5}$Clermont Universit\'{e}, Universit\'{e} Blaise Pascal, CNRS/IN2P3, LPC, Clermont-Ferrand, France\\
$ ^{6}$CPPM, Aix-Marseille Universit\'{e}, CNRS/IN2P3, Marseille, France\\
% ...
$ ^{11}$Physikalisches Institut, Ruprecht-Karls-Universit\"{a}t Heidelberg, Heidelberg, Germany\\
% ...
$ ^{33}$Universitat de Barcelona, Barcelona, Spain\\
$ ^{34}$Universidad de Santiago de Compostela, Santiago de Compostela, Spain\\
$ ^{35}$European Organization for Nuclear Research (CERN), Geneva, Switzerland\\
% ...
$ ^{38}$Nikhef National Institute for Subatomic Physics, Amsterdam, The Netherlands\\
% ...
$ ^{43}$H.H. Wills Physics Laboratory, University of Bristol, Bristol, United Kingdom\\
% ...
$ ^{49}$Oliver Lodge Laboratory, University of Liverpool, Liverpool, United Kingdom\\
% ...
$ ^{n}$LIFAELS, La Salle, Universitat Ramon Llull, Barcelona, Spain\\
% ...
}
\end{flushleft}

\end{titlepage}
\renewcommand{\thefootnote}{\arabic{footnote}}
\setcounter{footnote}{0}

\pagestyle{empty}  


\mbox{~}





\pagestyle{plain} 
\setcounter{page}{1}
\pagenumbering{arabic}


%

\section{Introduction}
\begin{figure}[hbt]
\begin{center}
\includegraphics[width=2 in]{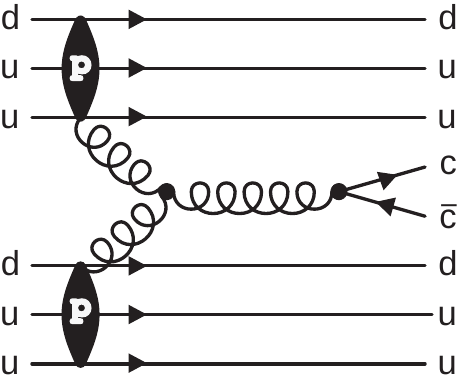}
\end{center}
\vspace{-3mm}
\caption{Production of $c\overline{c}$ quark pairs in a $pp$ collision via gluons.} \label{Ds-thy}
\end{figure}

Production of charm and bottom hadrons at the LHC in 7\,TeV $pp$ collisions is quite prolific. The bottom cross-section in the pseudorapidity region between 2 and 6 is about 80\,\mub \cite{Aaij:2010gn}, and the charm cross-section is about 30 times higher \cite{LHCb-CONF-2010-013}. In $pp$ collisions the production rates of charm and anti-charm particles need not be the same. While production diagrams are flavour symmetric, the hadronization process may prefer antiparticles to particles or vice versa.  Figure~\ref{Ds-thy} gives an example of $c\overline{c}$ production via gluon fusion. 
If the quarks that contribute to charm meson production are created in an independent fragmentation process, equal numbers of $D$ and $\overline{D}$ will be produced. On the other hand, if they combine with valence quarks in beam protons, the $\overline{c}$-quark can form a meson, while the $c$-quark can form a charmed baryon. Therefore, we may expect a small excess of $D_s^-$ over $D_s^+$ mesons. 
However, there are other subtle QCD effects that might contribute to a charm meson production asymmetry \cite{Chaichian:1993rh,Norrbin:2000jy,*Norrbin:2000zc}; we note for $b$ quarks the asymmetries are estimated to be at the 1\% level \cite{Aaij:2012qe}, and  we would expect them to be smaller for $c$ quarks, although quantitative predictions are difficult. Another conjecture is that any asymmetries might be reduced as particles are produced at more central rapidities. 
 
Measurements of \CP violating asymmetries in charm and bottom decays are of prime importance. 
These can be determined at the LHC if production and detection asymmetries are known.
The measurement of asymmetries in flavour specific modes usually involves detection of charged hadrons, and thus 
requires the relative detection efficiencies of $\pip$ versus $\pim$ or $\Kp$ versus $\Km$ to be determined.
While certain asymmetry differences can be determined by cancelling the detector response differences to positively and negatively charged hadrons \cite{Aaij:2011in}, more \CP violating modes can be measured if the relative detection efficiencies can be determined.

In this Letter we measure the production asymmetry, 
\begin{equation}
A_{\rm P}=\frac{\sigma(\Dsp)-\sigma(\Dsm)}{\sigma(\Dsp)+\sigma(\Dsm)},
\end{equation}
where $\sigma(\Dsm)$ is the inclusive prompt production cross-section.
We use $\Dspm\to\phi\pipm$ decays, where $\phi\to K^+K^-$. Since $\Dspm\to\phi\pipm$ is Cabibbo favoured, no significant \CP asymmetry is expected  \cite{Grossman:2006jg,Bergmann:1999pm}.  Assuming it to be vanishing, $A_{\rm P}$ is determined after correcting for the relative $D_s^+$ and $D_s^-$ detection efficiencies. Since the final
 states are symmetric in kaon production, this requires only knowledge of the relative $\pi^+$ and $\pi^-$ detection efficiencies, $\epsilon(\pi^+)/\epsilon(\pi^-)$.  

\section{Data sample and detector}

The data sample is obtained from $1.0\;\text{fb}^{-1}$ of integrated luminosity, collected with the \lhcb detector  \cite{Alves:2008zz} using $pp$ collisions at a center-of-mass energy of 7 TeV. 
The detector is a single-arm forward
spectrometer covering the pseudorapidity range $2<\eta <5$, designed
for the study of particles containing \bquark or \cquark quarks. The
detector includes a high precision tracking system consisting of a
silicon-strip vertex detector surrounding the $pp$ interaction region,
a large-area silicon-strip detector located upstream of a dipole
magnet with a bending power of about $4{\rm\,Tm}$, and three stations
of silicon-strip detectors and straw drift-tubes placed
downstream. The combined tracking system has a momentum resolution
$\Delta p/p$ that varies from 0.4\% at 5\gev to 0.6\% at 100\gev.\footnote{We work in units with $c$=1.} 
Charged hadrons are identified using two
ring-imaging Cherenkov (RICH) detectors. Photon, electron and hadron
candidates are identified by a calorimeter system consisting of
scintillating-pad and pre-shower detectors, an electromagnetic
calorimeter and a hadronic calorimeter. Muons are identified by a muon
system composed of alternating layers of iron and multiwire
proportional chambers. The trigger consists of a hardware stage, based
on information from the calorimeter and muon systems, followed by a
software stage which applies a full event reconstruction. Approximately 40\% of the data was taken with the magnetic field directed away from the Earth (up) and the rest down. We exploit the fact that certain detection asymmetries cancel if data from different magnet polarities are combined.

Events are triggered by the presence of a charm hadron decay. The hardware trigger requires at least one hadronic transverse energy deposit of approximately 3 GeV. Subsequent software triggers and selection criteria require a subset of tracks to not point to a primary $pp$ collision vertex (PV), and form a common vertex. 

\section{Measurement of relative pion detection efficiency}
In order to measure $\epsilon(\pi^+)/\epsilon(\pi^-)$, we use the decay sequence $D^{*+}\to\pi^+_s D^0$, $D^0\to K^-\pi^+\pi^+\pi^-$, and its charge-conjugate decay, where $\pi^+_s$ indicates the ``slow" pion coming directly from the $D^{*+}$ decay.
Assuming that the $D^{*\pm}$ comes from the PV, there are sufficient kinematic constraints to detect this decay even if one pion from the $D^0$ decay is missed. We call these ``partially" reconstructed decays.  We can also ``fully" reconstruct this decay. The ratio of fully to partially reconstructed decays provides a measurement of the pion reconstruction efficiency. We examine $D^{*+}$ and $D^{*-}$ candidate decays separately, and magnet up data separately from magnet down data. The latter is done to test for any possible left-right detector asymmetries. In both cases the missing pion's charge is required to be opposite of that of the detected kaon.

Kaon and pion candidates from candidate $D^0$ decays are required to have transverse momentum, $\pt>400$\,MeV, and a track quality fit with $\chi^2$ per number of degrees of freedom (ndf)\,$<3$, keeping more than 99\% of the good tracks.
The distance of closest approach of track candidates to the PV is called the impact parameter (IP).  A restrictive requirement is imposed on  the  IP $\chi^2$, which measures whether the track is consistent with coming from the PV, to be greater than 4. In addition both particles must be identified in the RICH. 
For the $\pi^+_s$, the \pt requirement is lowered to 250 MeV, with both IP $<$~0.3\,mm and IP $\chi^2\,<\,4$ being required. Further tight restrictions are placed on $D^0$ candidates. The candidate tracks from the $D^0$ decay must fit to a common vertex with $\chi^2$/ndf$\,<\,$6, the $D^0$ candidates must have a flight distance of at least 4 mm from the PV and have a flight distance $\chi^2>\,$120.  We require $1.4\,<\,m(K^-\pi^+\pi^-)\,<\,1.7$\,GeV, and that the invariant mass of the $\pi^+\pi^-$ candidates must be within $\pm$200 MeV of the $\rho(770)$ mass, to improve the signal to background ratio.

\begin{figure}[!b]
\begin{center}
\includegraphics[width=6 in]{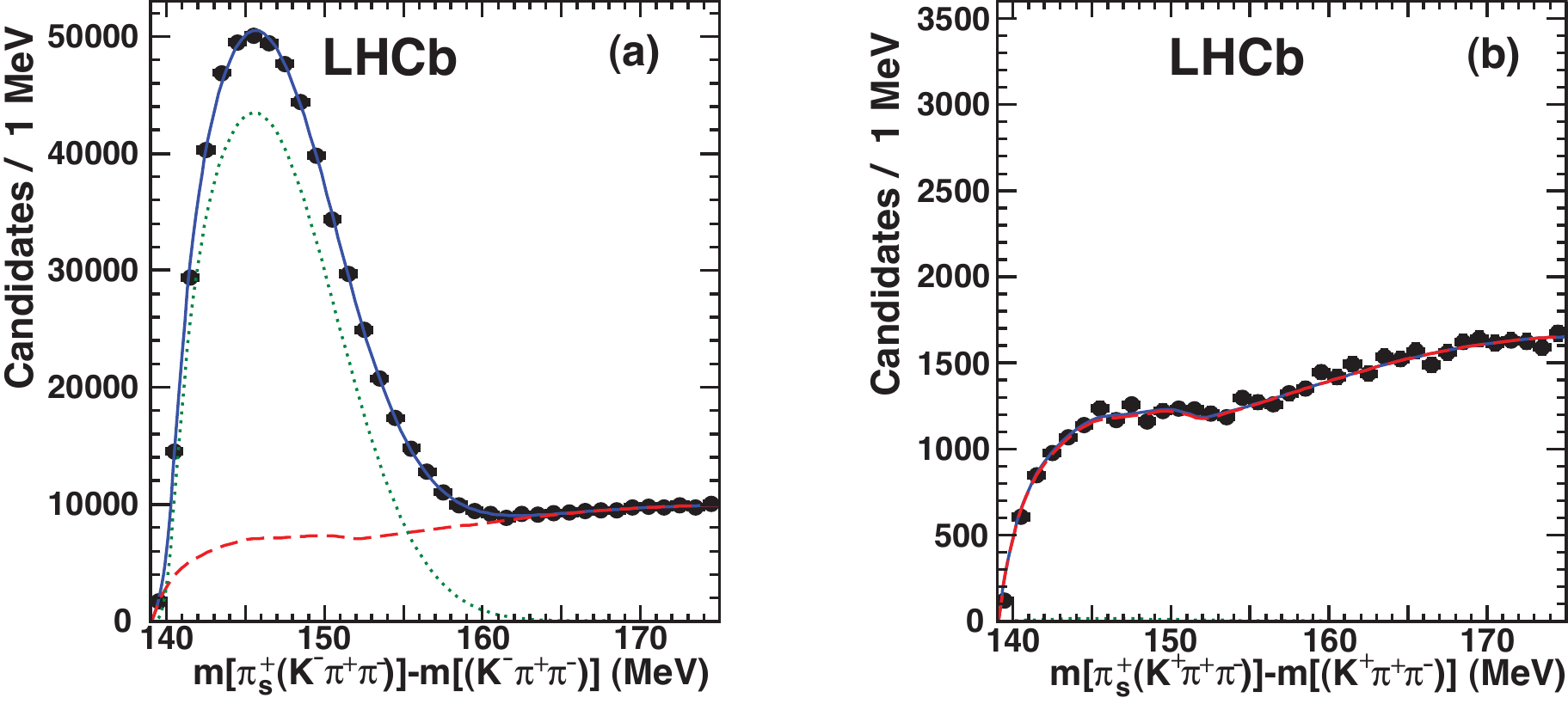}
\end{center}
\vspace{-6 mm}
\caption{Distributions of  mass differences in partial reconstruction for (a) RS $m(\pip_{s}\Km\pipi)-m(\Km\pipi)$ and (b) WS $m(\pip_{s}\Kp\pipi)-m(\Kp\pipi)$ candidates, for magnet up data. The (green) dotted line shows the signal, the (red) dashed line the background, and the (blue) solid line the total. The fit shapes are defined in Appendix A.
 } \label{mass difference part piplus}
\end{figure}

We select partially reconstructed  right-sign (RS) $D^0$ candidates by examining the mass difference $\Delta m_{\text{prt}} = m(\pip_{s}\Km\pipi)-m(\Km\pipi)$. Wrong-sign (WS) candidates are similarly selected but by requiring that the charge of the kaon be the same as that of the $\pi_s^+$. 
Figure~\ref{mass difference part piplus} shows distributions of  $\Delta m_{\text{prt}}$ for magnet up data. Note that the yield of WS events is reduced due to a prescale factor applied in the selection. 

In order to determine the size of the signals above the background we perform simultaneous binned maximum likelihood fits to the RS and WS distributions. The parametrization of the signal probability density function (PDF) is given in Appendix A.
The signal and background PDFs are identical for RS and WS \Dz and \Dzb events, only the absolute normalizations are allowed to differ. We also include a ``signal" term in the fit to WS events to account for the doubly-Cabibbo-suppressed (DCS) signals. The ratio of the DCS signal in WS events to the signal in RS events is fixed to that obtained in the mass difference fit in full reconstruction.

Using momentum and energy conservation and knowledge of the direction of the $D^0$ flight direction,
the inferred three-momentum of the missing pion, $\vec{P}_{\!\rm inf}$, is reconstructed using a kinematic fitting technique \cite{LagrangeM}. Our resolution on inferred pions may be determined from the the fully reconstructed \Dstarpm sample, by removing one detected pion whose three-momentum is well known, and treating the track combination as if it was partially reconstructed.
We then have both detected and inferred momentum, and thus a measurement of the missing pion momentum resolution distribution, $\Delta P / P=(P_{\text{detected}}-P_{\text{inf}})/P_{\text{detected}}$.
For further study, we take only combinations with good inferred resolution by accepting those where ${P}_{\!\rm inf}$ divided by its calculated uncertainty is greater than two and also where the transverse component of ${P}_{\!\rm inf}$  divided by its uncertainty is greater than 2.5; this eliminates about 37\% of the sample.

In our sample of partially reconstructed events, we subsequently look for fully reconstructed decays by searching for the missing track. 
Candidate tracks must have $p>2$\,GeV, $\pt>300$\,MeV and be identified as a pion in the RICH.
They also must form a vertex with the other three tracks from the decay with a vertex fit $\chi^2$/ndf$<6$, and have a four-track invariant mass within 30 MeV of the $D^0$ mass peak. Certain areas of the detector near its edges preferentially find only one charge or the other depending on magnet polarity.  We remove fully reconstructed candidates where the detected pion projects to these regions, discarding 3\% of the candidates.

\begin{figure}[hbt]
\begin{center}
\includegraphics[width=5.9 in]{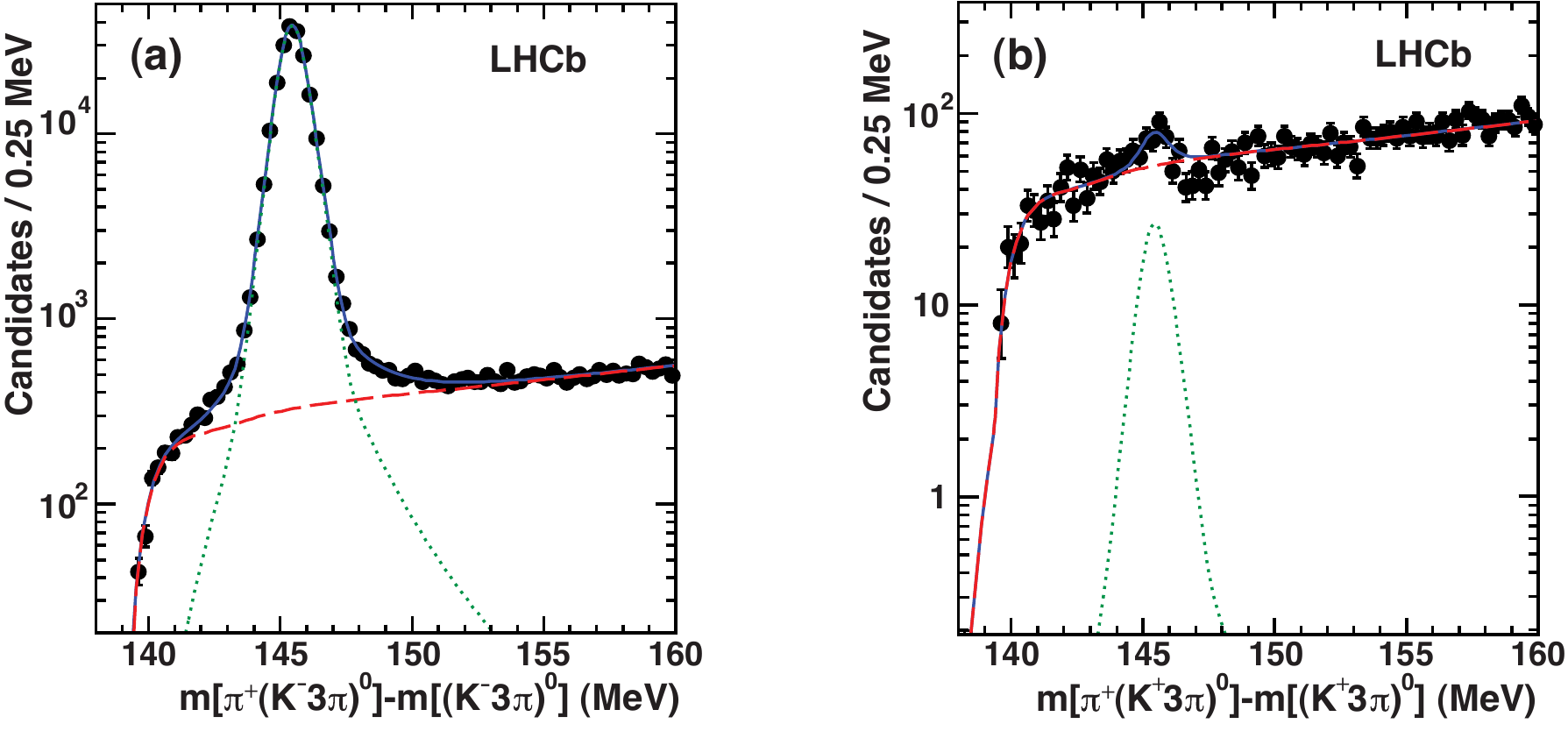}
\end{center}
\vspace{-4mm}
\caption{Distributions of the mass difference $\Delta m_{\text{full}}$  for (a) RS  and (b) WS  events using magnet up data. The (green) dotted line shows the signal, the (red) dashed line the background, and the (blue) solid line the total. The fit shapes are defined in Appendix B.
 } \label{mass difference full piplus}
\end{figure}

The mass difference for fully reconstructed combinations, $\Delta m_{\text{full}} = m(\pip_{s}\Km\pipi\pip)-m(\Km\pipi\pip)$ is shown in Fig.~\ref{mass difference full piplus}, for both RS and WS cases. Only $D^{*+}$ data in the magnet up configuration are shown. The  shape of the mass difference signal PDF is described in Appendix B.

\begin{figure}[!hbt]
\begin{center}
\includegraphics[width=5.8 in]{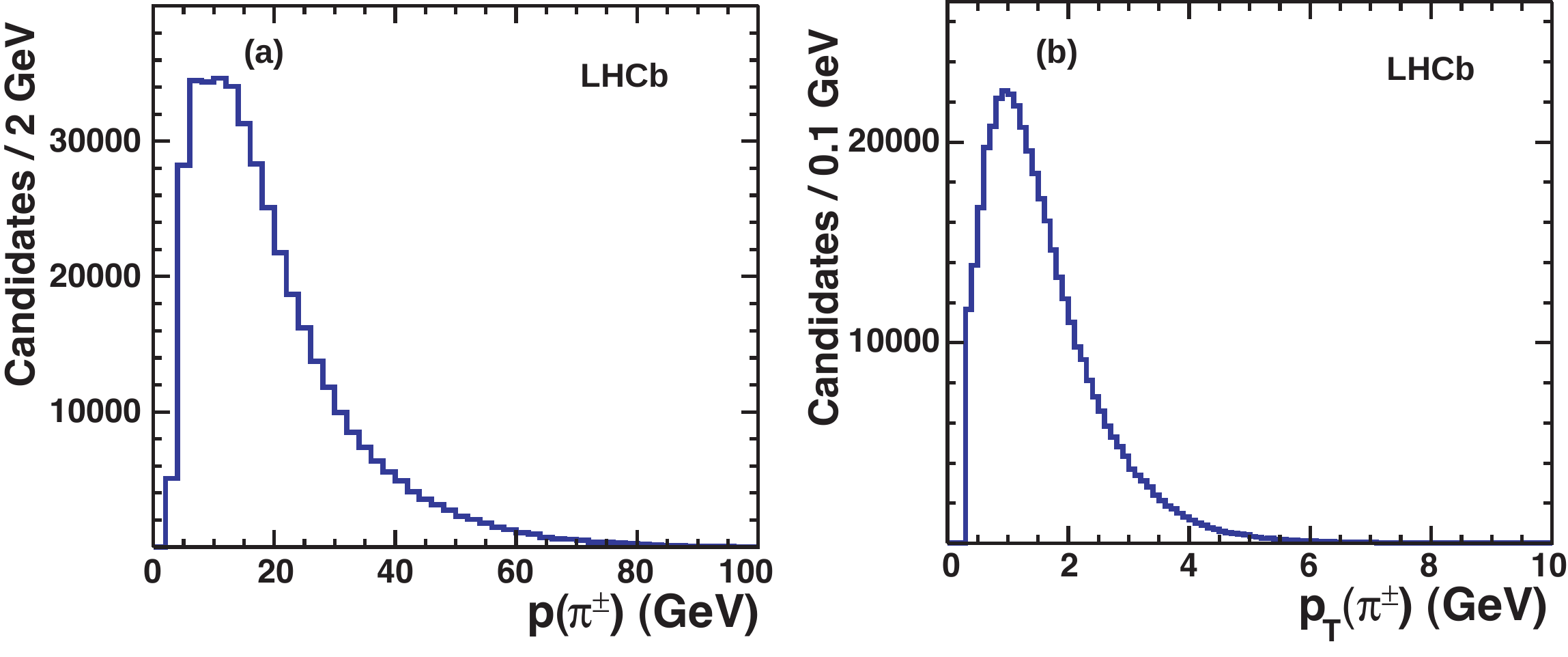} \end{center}
\vspace{-8mm}
\caption{Distribution of fully reconstructed signal candidates for magnet up data as a function of pion  (a)  $p$ and (b) \pt.} \label{full_yields}
\end{figure}

In order to extract the signal yields, we perform a binned maximum likelihood fit to the \Dstarp and \Dstarm events, both RS and WS, simultaneously. Table~\ref{tab:summary of efficiency ratio magupdown} lists the signal yields for partial reconstruction, $N_{\rm prt}$, and full reconstruction, $N_{\rm full}$. The efficiency ratios are derived from ratios of RS yields,
 $\epsilon(\pip)=N_{\text{full}}(\Dz\pip_{s})/N_{\text{prt}}(\Dz\pip_{s})$ and $\epsilon(\pim)=N_{\text{full}}(\Dzb\pim_{s})/N_{\text{prt}}(\Dzb\pim_{s})$. (The absolute efficiency inferred from these yields includes geometric acceptance effects.) The $p$ and \pt spectra of the $\pi^{\pm}$ used for the efficiency measurement are shown in Fig~\ref{full_yields}.
\begin{table}[htb]
  \caption{Event yields for partial and full reconstruction.}
  \begin{center}\begin{tabular}{ccc} \hline
   
     Category &{Magnet up} & Magnet down  \\
     	\hline
     	$N_{\text{prt}}(\Dz\pip_{s})$            & 460,005$\pm$890 &671,638$\pm$1020  \\
      $N_{\text{prt}}(\Dzb\pim_{s})$           & 481,823$\pm$873 & 694,268$\pm$1035 \\
      $N_{\text{full}}(\Dz\pip_{s})$            & 207,504$\pm$465 & 299,629$\pm$~570\\
      $N_{\text{full}}(\Dzb\pim_{s})$           & 219,230$\pm$478 & 308,344$\pm$~579\\
      \hline

\end{tabular}\end{center}
  \label{tab:summary of efficiency ratio magupdown}
\end{table}

\begin{figure}[hbt]
\begin{center}
\vspace{-.1mm}
\includegraphics[width=3.9 in]{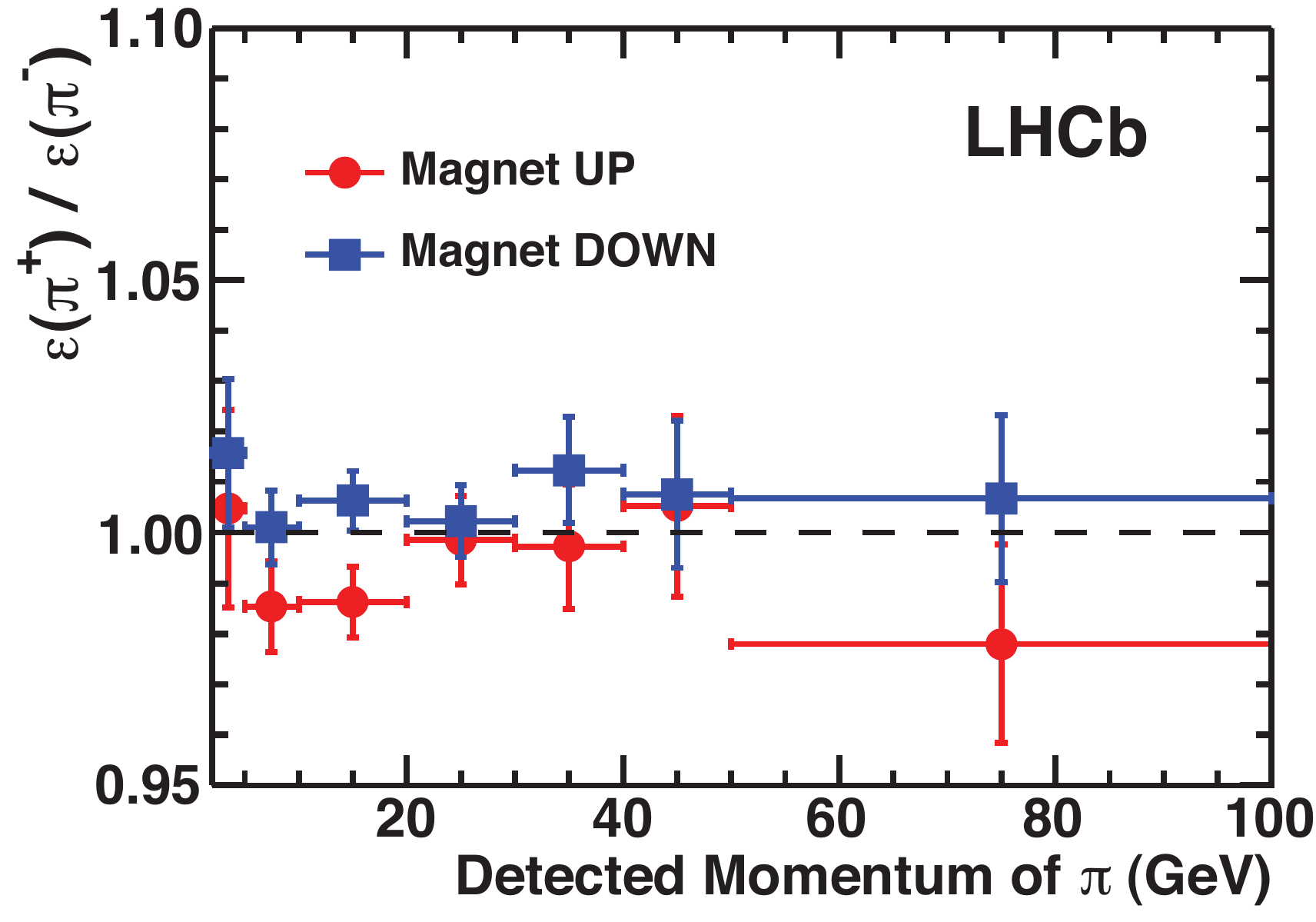}
\end{center}
\vspace{-8mm}
\caption{Relative detection efficiency in bins of detected pion momentum:  (red) circles represent data taken with magnet polarity up and (blue) squares show data taken with magnet polarity down. Only statistical errors are shown.} \label{relative_efficiency}
\end{figure}

The ratios of pion detection efficiencies are 0.9914$\pm$0.0040 and 1.0045$\pm$0.0034 for magnet up and magnet down, respectively, with statistical uncertainties only. To obtain the efficiency ratio as a function of momentum 
 we need to use the inferred momentum of the missing pion. Because of finite resolution it needs to be corrected. 
This is accomplished through an unfolding matrix estimated using the fully reconstructed sample by comparing the measured momentum of a found pion that is then ignored and its momentum inferred using the kinematic fit. 
The efficiency ratio is shown as a function of momentum in Fig.~\ref{relative_efficiency}. Most systematic uncertainties cancel in the efficiency ratio, however, some small residual effects remain. To assess them we change the signal and background PDFs in full and partial reconstruction by eliminating, in turn, each of the small correction terms to the main functions. The full fit is then repeated. Each change in the efficiency ratios are between 0.01$-$0.02\%. We also change the amount of DCS decays by the measured uncertainty in the branching fraction. This also gives a 0.020\% change. The total systematic error is 0.045\%.  Furthermore, the entire procedure was checked using simulation.

\begin{figure}[hbt]
\begin{center}
\includegraphics[width=4 in]{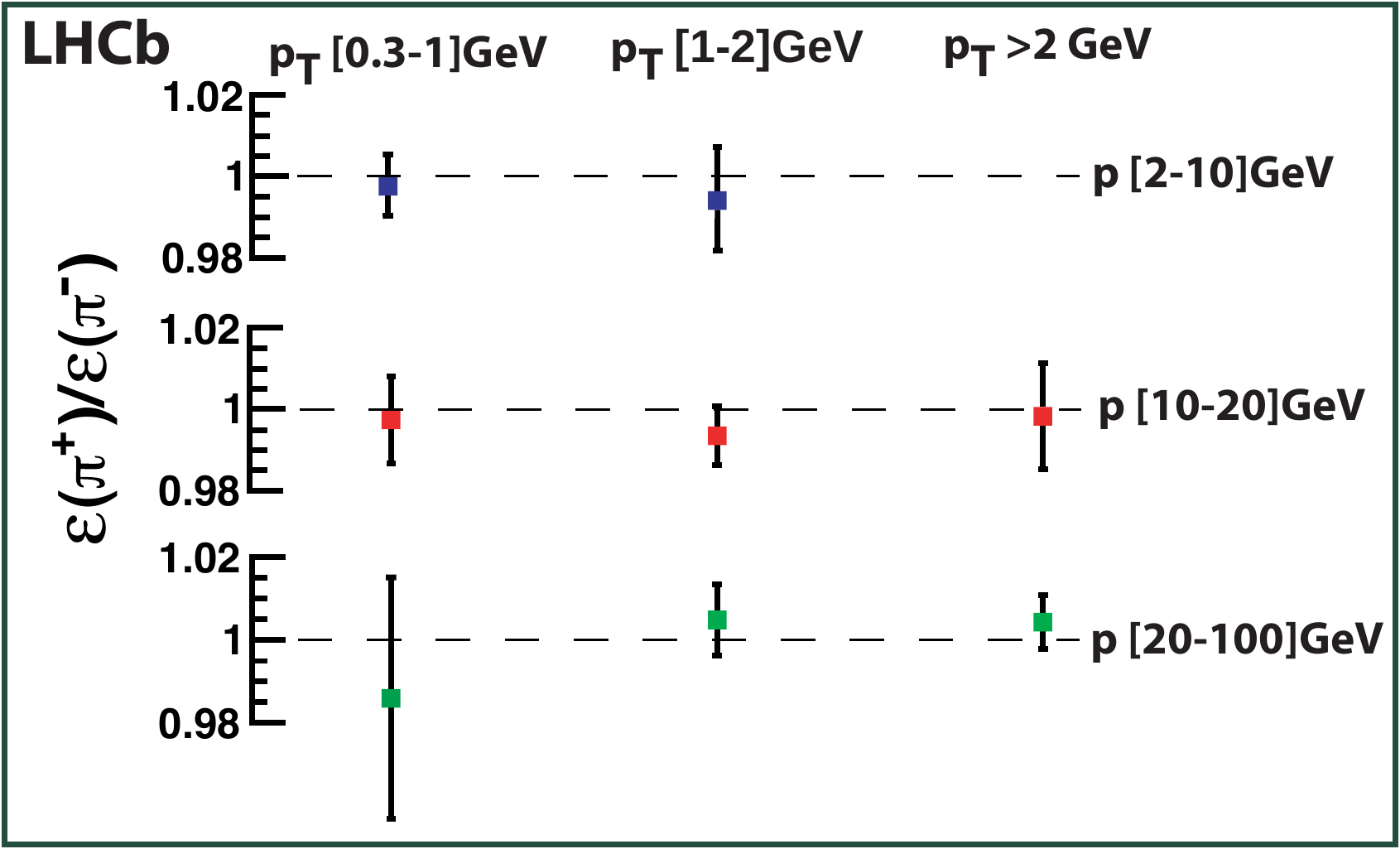}
\end{center}
\vspace{-6mm}
\caption{Relative efficiency averaged over magnet up and magnet down samples versus pion $p$ and $p_{T}$.} \label{p_slice_all}
\end{figure}

Although we correct relative pion efficiencies as a function of \ptot, it is possible that there also is a \pt dependence that would have an effect if the \pt distributions of the $D^{*\pm}$ and $D_s^{\pm}$ were different.
The efficiency ratios for different slices are shown in Fig.~\ref{p_slice_all}. For a fixed \ptot interval there is no visible \pt dependence.

\begin{figure}[hbt]
\begin{center}
\includegraphics[width=6 in]{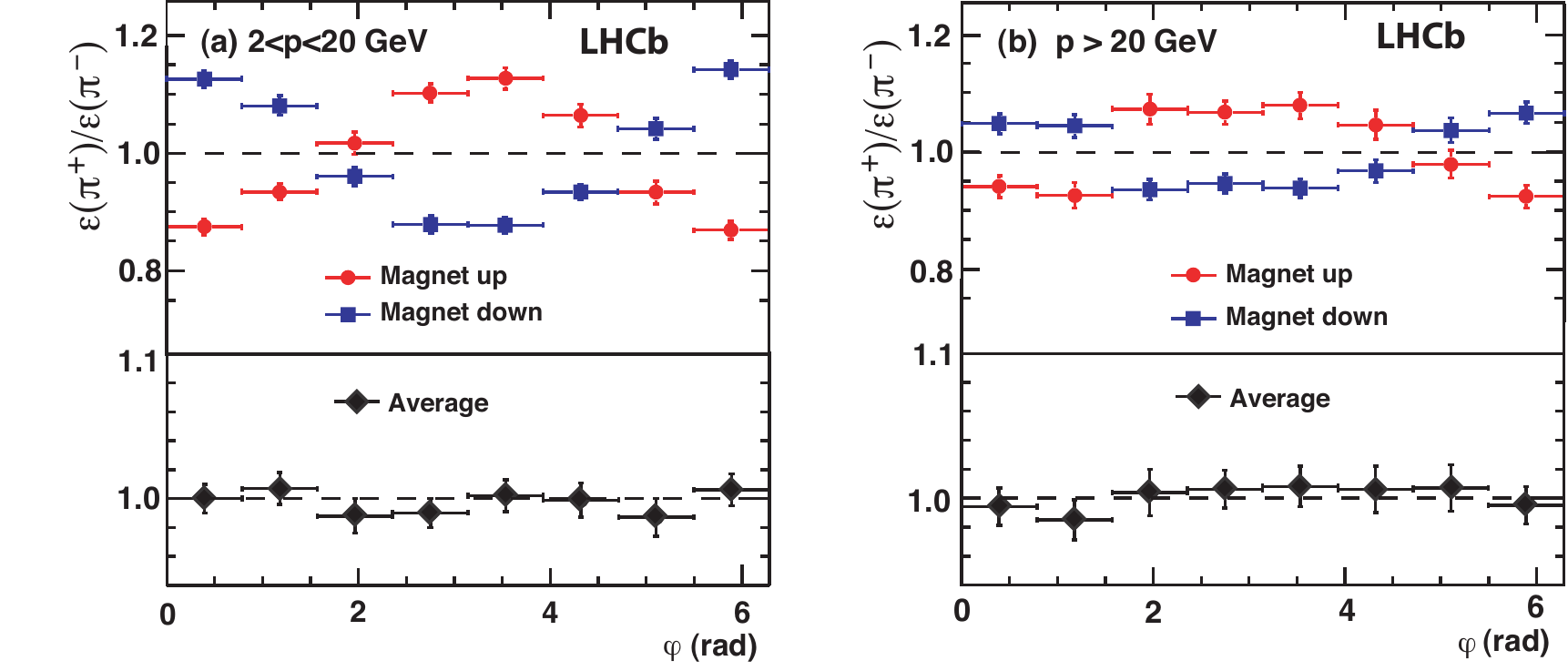}
\end{center}
\vspace{-15pt}
\caption{Azimuthal angle distribution of $\epsilon(\pi^+)/\epsilon(\pi^-)$ for magnet up data (red circles) and magnet down data (blue squares), and their average
(black diamonds) for (a) pion momentum $2<p<20$ GeV and (b) $p>20$ GeV.} \label{efficiency_ratio_vs_Phi}
\end{figure}

The relative pion efficiencies are consistent with being independent of $p$ and \pt. The tracking acceptance does depend, however, on the azimuthal production angle of the particles, $\varphi$. This is mostly because tracks can be swept into the beam pipe and not be detected by the downstream tracking system. Therefore, for purposes of the production asymmetry analysis we determine $\epsilon(\pi^+)/\epsilon(\pi^-)$ as a function of $\varphi$ in two momentum intervals: $2-20$ GeV, and above 20 GeV. The r.m.s. resolution on the inferred $\varphi$ is 0.25\,rad, much smaller than the $\pi$/4 bin size. The correction factors are shown in Fig.~\ref{efficiency_ratio_vs_Phi}. The average correction for magnet up and magnet down is consistent with unity. Thus any residual biases in the $D_s^{\pm}$ yields due to $\pi^+/\pi^-$ asymmetries will also cancel in the average.

\section{\boldmath \Dspm production asymmetry}
The decay $D_s^{\pm}\to K^+K^-\pi^{\pm}$ is used with the invariant mass of the $K^+K^-$ required to be within $\pm$20\,MeV of the $\phi$ mass.
Events are triggered at the hardware level by requiring that either the $K^+$ or the $K^-$ deposits more than 3 GeV of transverse energy in the hadron calorimeter. Subsequent software triggers are required to select both $\phi$ decay products.

To select a relatively pure sample of $D_s^{\pm}\to K^+K^-\pi^{\pm}$ candidates each track is required to have $\chi^2$/ndf $<$4, $\pt>\,$300\,MeV, IP $\chi^2>4$, and be identified in the RICH.  All three candidate tracks from the $D_s^{\pm}$ have $\pt>2$\,GeV, and must form a common vertex that is detached from the PV. The $\chi^2$ requiring all three tracks to come from a common origin must be $<8.33$, this decay point must be at least 100\,$\mu$m from the PV, and the significance of the detachment must be at least 10 standard deviations.  The $D_s^{\pm}$ candidates' momentum vector must also point to the PV, which reduces contamination from $b$-hadron decays to the few percent level. We remove signal candidates with pions which pass through the detector areas with large inherent asymmetries, as we did to measure the relative pion efficiencies. 

\begin{figure}[t]
\begin{center}
\includegraphics[width=5.9 in]{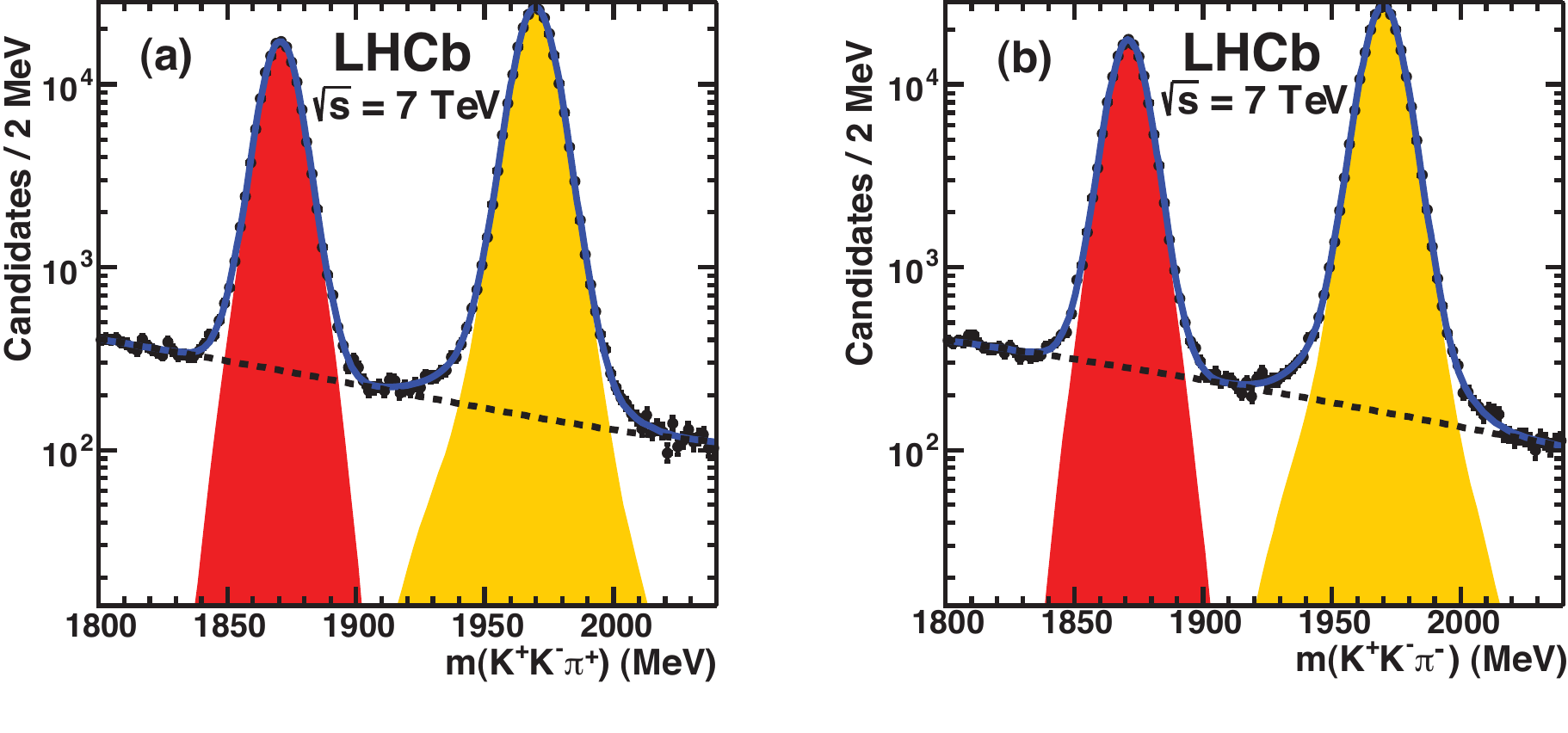}
\end{center}
\vspace{-30pt}
\caption{Invariant mass distributions for (a) $\KpKm\pip$ and (b) $\Km\Kp\pim$ candidates, when $m(K^+K^-)$  is within $\pm$20 MeV of the $\phi$ mass, for the dataset taken with magnet polarity down. The shaded areas represent signal, the dashed line the background and the solid curve the total. \label{Ds mass overall fit magdown}}
\end{figure}

Figure~\ref{Ds mass overall fit magdown} shows the invariant mass distributions for (a) $\KpKm\pip$ and (b) $K^+K^-\pi^-$ candidates for data taken with magnet polarity down. We perform a binned maximum likelihood fit to extract the signal yields. The fitting functions for both $D^{\pm}$ and  $D_s^{\pm}$ signals are triple Gaussians where all parameters are allowed to vary, except two of the Gaussians are required to have the same mean. The background function is a second order polynomial. The numbers of $D_s^{\pm}$ events obtained from the fits are listed in Table~\ref{tab:Ds}.

\begin{table}[b]
  \caption{Fitted numbers of $D_s^{\pm}$ events for both magnet up and down data.}
  \begin{center}\begin{tabular}{lcc}\hline

&{Magnet up}&{Magnet down}  \\\hline
$D_s^+$ &152,696$\pm$448&  230,860$\pm$514  \\
$D_s^-$ &154,209$\pm$438 & 233,266$\pm$549 \\
\hline

  \end{tabular}\end{center}
  \label{tab:Ds}
\end{table}

\begin{figure}[t]
\begin{center}
\includegraphics[width=5.9 in]{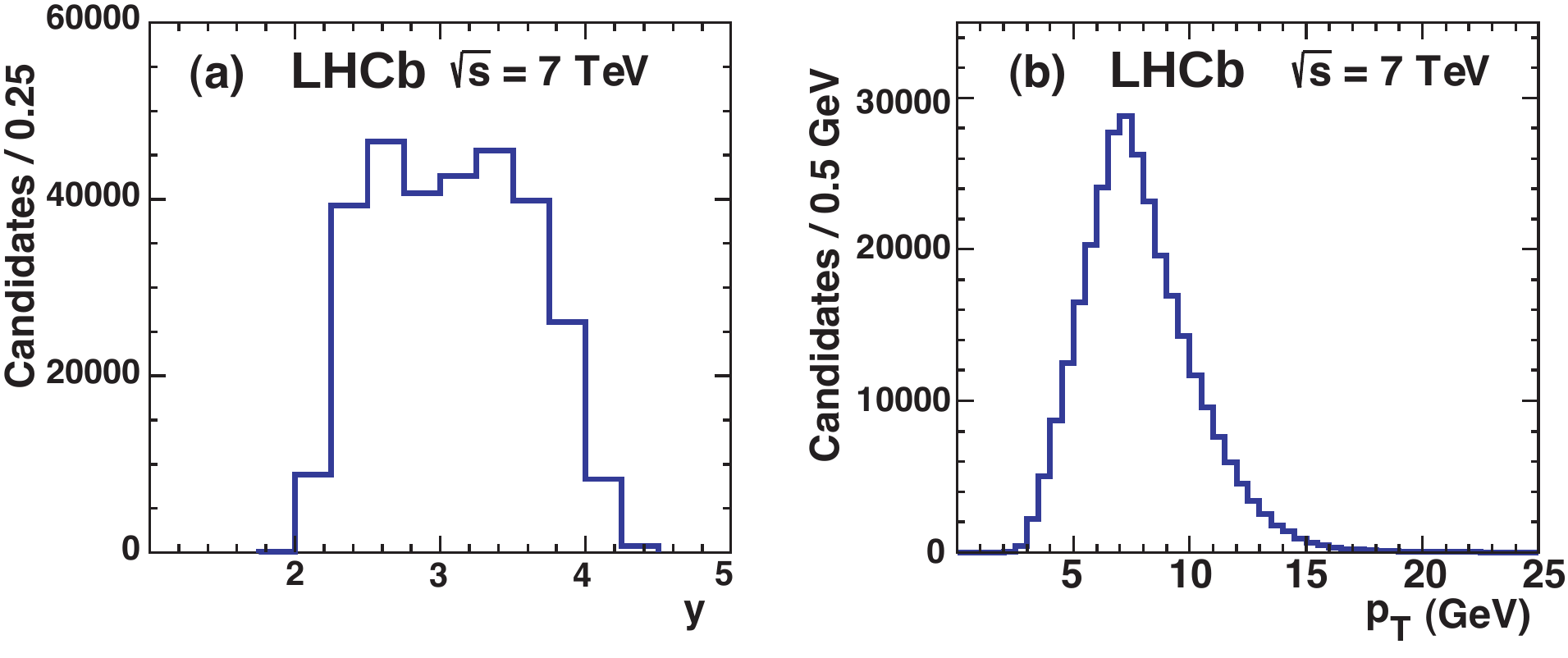}
\end{center}
\vspace{-30pt}
\caption{ (a) \Dspm rapidity distribution (b) \Dspm $p_{T}$ distribution for background subtracted magnet up data. The
statistical uncertainty on the number of events in each bin is smaller than the line thickness.} \label{fig:Ds_Pt_Rapidity_spectra}
\end{figure}

The rapidity of the \Ds is defined as
\begin{equation}
y=\frac{1}{2}\ln\frac{E+p_{z}}{E-p_{z}},
\end{equation} where $E$ and $p_{z}$ are the energy and $z$ component of the \Dspm momentum. We  measure the production asymmetry $A_{\rm P}$ as a function of both \Dspm $y$ and \pt. In each $y$ or \pt bin we extract the efficiency corrected ratio of yields by applying corrections as a function of azimuthal angle in the two pion momentum intervals defined previously. Magnet up and down data are treated separately. 
The $y$ and \pt distributions are shown in Fig.~\ref{fig:Ds_Pt_Rapidity_spectra}. Here sidebands in $KK\pi$ mass have been used to subtract the background, where the sidebands are defined as between $30-70$\, MeV above and below the peak mass value of 1969\,MeV. As this interval is twice as wide as the signal peak, we weight these events by a factor of 1/2.

\begin{figure}[t]
\begin{center}
\includegraphics[width=6 in]{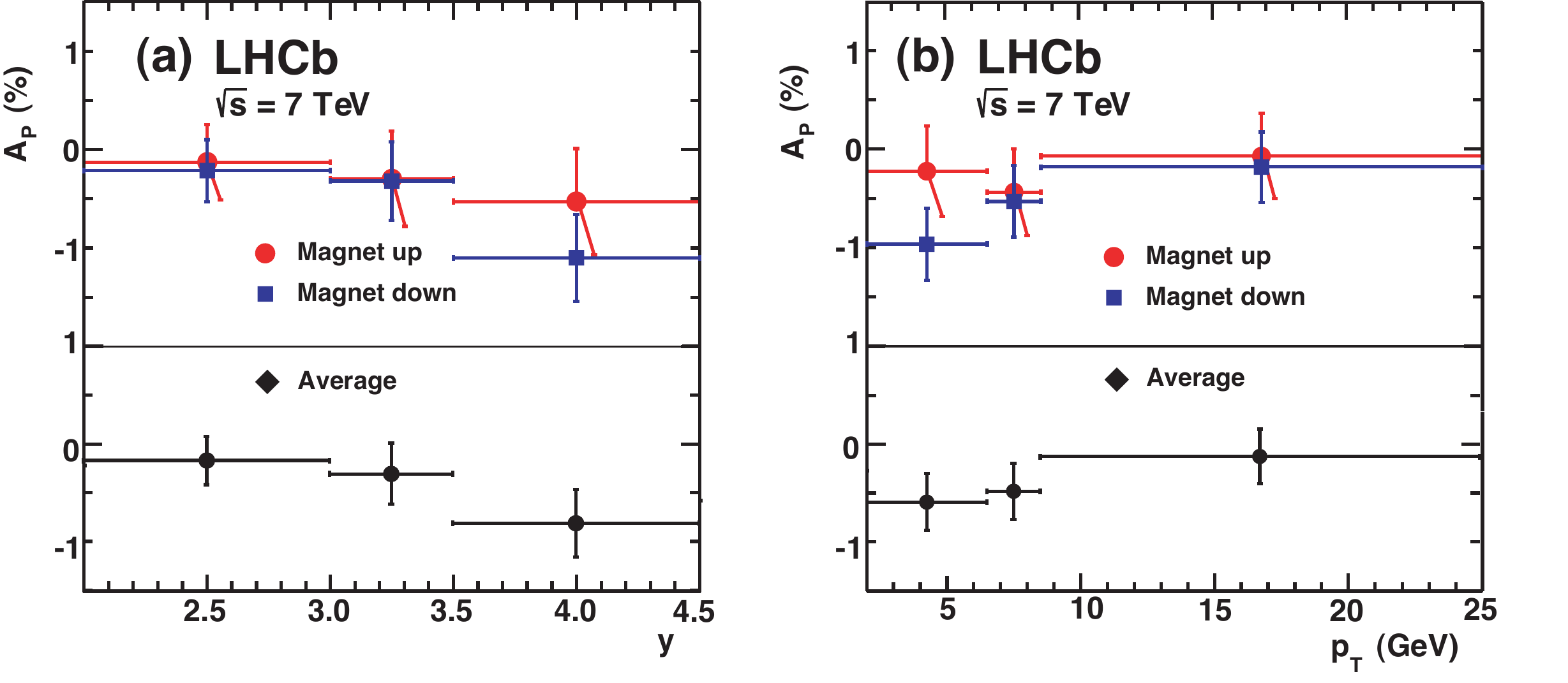}
\end{center}
\vspace{-20pt}
\caption{Observed production asymmetry $A_{\rm P}$ as a function of (a) y, and (b) \pt. The errors shown are statistical only. } \label{Ap-y-pt}
\end{figure}

Figure~\ref{Ap-y-pt} shows $A_{\rm P}$ as a function of either $y$ or \pt. The error bars reflect only the statistical uncertainties, which includes both the statistical errors on $\epsilon(\pi^+)/\epsilon(\pi^-)$ and the $D^{\pm}_s$ yields; the error bars are partially correlated, the uncertainties from the $D_s^{\pm}$ yields are about half the size of those shown.  The values in  \pt and $y$ intervals are listed in Table~\ref{tab:Ap}.
\begin{table}[b]
  \caption{$A_{\rm P}$ (\%) shown as a function of both $y$ and \pt.}
  \begin{center}
\begin{tabular}{cccc}\hline
 \pt (GeV) & \multicolumn{3}{c}{~~$y$}\\
& ~\,\,$2.0-3.0$ & ~\,\,$3.0-3.5$ & ~\,\,$3.5-4.5$\\\hline
 $2.0-~\,6.5$ &~~$0.2\pm0.5$ & $-0.7\pm 0.5$ & $-0.9\pm 0.4$\\
 $6.5-~\,8.5$ & $-0.3\pm0.4$ & $~~\,0.1\pm 0.5$ & $-1.2\pm 0.5$\\
 $8.5-25.0$ & ~~$0.2\pm0.3$ & $-0.3\pm 0.5$ & $-1.0\pm 0.8$\\
\hline
\end{tabular}\end{center}
  \label{tab:Ap}
\end{table}
An average asymmetry in this $y$ and \pt region can be derived by weighting the asymmetry in each bin by the
production yields. Thus we take the asymmetry in each $y$ and \pt interval, weight by the measured event yields divided by the reconstruction efficiencies. 
The resulting integrated production asymmetry $A_{\rm P}$ is
$(-0.20\pm0.34)$\%, and $(-0.45\pm0.28)$\%,
for magnet up and magnet down samples, respectively. The errors are statistical only.
Averaging the two results, giving equal weight to each to cancel any residual systematic biases, gives
\begin{equation*}
    A_{\rm P}=(-0.33\pm0.13\pm0.18\pm0.10)\%,
\end{equation*}
where the first uncertainty is statistical from the $D_s^{\pm}$ yields, the second statistical due to the error on the efficiency ratio and the third systematic. 
The systematic uncertainty on $A_{\rm P}$ has several contributions. Uncertainties due the background shape in the \Dspm mass fit are evaluated using a higher order polynomial function, that gives a $0.06\%$ change. Statistical uncertainty on MC efficiency adds 0.06\%.  Constraining the signal shapes of the $D_s^+$ and $D_s^-$ to be the same makes a 0.04\% difference. Possible changes in detector acceptance during magnet up and magnet down data taking periods are estimated to contribute 0.03\%. The systematic uncertainty from the pion efficiency ratio contributes 0.02\%. Differences in the momentum distributions of $K^-$ and $K^+$ that arise from interference with an S-wave component under the $\phi$ peak can introduce a false asymmetry \cite{Staric:2011en}.  For our relatively high momentum $D_s^{\pm}$ mesons this is a 0.02\% effect. Contamination from $b$ decays causes a negligible effect. Adding all sources in quadrature, the overall systematic uncertainty on $A_{\rm P}$ is estimated to be $0.10\%$.

\section{Conclusions}
We have developed a method using partially and fully reconstructed $\Dstarpm$ decays to measure the relative detection efficiencies of positively and negatively charged pions as a function of momentum.  Applying this method to $D_s^{\pm}$ mesons produced directly in $pp$ collisions, \emph{i.e.} not including those from decays of $b$ hadrons, we measure the overall production asymmetry in the rapidity region 2.0 to 4.5, and $\pt>2$\,GeV as
\begin{equation}
A_{\rm P}=\frac{\sigma(\Dsp)-\sigma(\Dsm)}{\sigma(\Dsp)+\sigma(\Dsm)}=(-0.33\pm0.22\pm0.10)\%.
\end{equation}
The asymmetry is consistent with being independent of \pt, and also consistent with being independent of $y$, although there is a trend towards smaller $A_{\rm P}$ values at more central rapidity. These measurements are consistent with theoretical expectations \cite{Chaichian:1993rh,Norrbin:2000jy,*Norrbin:2000zc}, provide significant constraints on models of $D_s^{\pm}$ production, and can be used as input for \CP violation measurements.

\section*{Acknowledgments}
\noindent We express our gratitude to our colleagues in the CERN accelerator
departments for the excellent performance of the LHC. We thank the
technical and administrative staff at CERN and at the LHCb institutes,
and acknowledge support from the National Agencies: CAPES, CNPq,
FAPERJ and FINEP (Brazil); CERN; NSFC (China); CNRS/IN2P3 (France);
BMBF, DFG, HGF and MPG (Germany); SFI (Ireland); INFN (Italy); FOM and
NWO (The Netherlands); SCSR (Poland); ANCS (Romania); MinES of Russia and
Rosatom (Russia); MICINN, XuntaGal and GENCAT (Spain); SNSF and SER
(Switzerland); NAS Ukraine (Ukraine); STFC (United Kingdom); NSF
(USA). We also acknowledge the support received from the ERC under FP7
and the Region Auvergne.

\section*{Appendix A: Fitting functions for partial reconstruction}
\label{App:partial}
The signal probability density function (PDF) is given by:
\begin{align}
f_{\rm sig} (\Delta m_{\text{prt}})&=f_{\rm eff}(\Delta m_{\text{prt}}) \cdot BG(\Delta m_{\text{prt}};\sigma_l,\sigma_r,\mu),~{\rm where}
\\
\nonumber
BG(\Delta m_{\text{prt}})&=
\begin{cases} 
\frac{\sigma_{l}}{\sigma_{l}+\sigma_{r}}G(\Delta m_{\text{prt}};\mu,\sigma_{l}) & \text{if $\Delta m_{\text{prt}} \leq \mu$,}
\\
\frac{\sigma_{r}}{\sigma_{l}+\sigma_{r}}G(\Delta m_{\text{prt}};\mu,\sigma_{r}) & \text{if $\Delta m_{\text{prt}} > \mu$.}
\end{cases}
\end{align} 
$G(\Delta m_{\text{prt}}$;$\mu$,$\sigma$) is a Gaussian function with mean $\mu$ and width $\sigma$, and  $BG$($\Delta m_{\text{prt}}$) is a bifurcated Gaussian function. The efficiency function $f_{\text{eff}}(\Delta m_{\text{prt}})$ is defined as:
\begin{equation}
f_{\text{eff}}(\Delta m_{\text{prt}})=
\begin{cases}
\frac{|a(\Delta m_{\text{prt}}-\Delta m_{0})|^{N}}{1+|a(\Delta m_{\text{prt}}-\Delta m_{0})|^{N}} & \text{if $\Delta m_{\text{prt}}-\Delta m_{0} \geq 0$,}
\\ 
0 & \text{if $\Delta m_{\text{prt}}-\Delta m_{0}<0$,} 
\end{cases} 
\end{equation}
where $a$, $N$, and $\Delta m_{0}$ are fit parameters.
The resolution function (the bifurcated Gaussian function) is multiplied by the efficiency function $f_{\text{eff}}(\Delta m_{\text{prt}})$ in order to account for the ``turn-off" behaviour of the quantity $\Delta m_{\text{prt}}$ near the threshold (pion mass). There are in total six shape parameters in this signal PDF which are left to vary in the fit.

The background PDF is taken as a threshold function with the  inclusion of extra components to obtain a good description of the WS combinations. 
It is defined similarly:
\begin{equation}
\label{RooDstD0BG}
f_{\rm bkg}(\Delta m_{\text{prt}}) = f^{*}(\Delta m_{\text{prt}}) \cdot (c_{2}\Delta m_{\text{prt}}^{2}+c_{1}\Delta m_{\text{prt}}+1)- f_{1} \cdot BG(\Delta m_{\text{prt}})+ f_{2}  \cdot G(\Delta m_{\text{prt}}), 
\end{equation}
\begin{equation}
\label{eq:fstar}
f^{*}(\Delta m_{\text{prt}})=\left[1-\exp(-(\Delta m_{\text{prt}}-\Delta m^{p}_{0})/c_{p})\right] \cdot a_{p}^{\Delta m_{\text{prt}}/\Delta m^{p}_{0}}+b_{p}(\Delta m_{\text{prt}}/\Delta m^{p}_{0}-1).
\end{equation} 
The parameters used in the background functions $BG$ and $G$ are different than the ones used in the signal functions.
There are in total 11 shape parameters in the background PDF that are determined by the fit. We also fit using $f^{*}(\Delta m_{\text{prt}})$ as the background PDF alone to estimate the systematic uncertainty on the efficiency ratio.

\section*{Appendix B: Fitting functions for full reconstruction}
The signal PDF is defined as:
\begin{align}
\nonumber
f_{\rm sig} (\Delta m_{\text{full}})&=f_{1}G(\Delta m_{\text{full}};\mu_{1},\sigma_{1})+f_{2}G(\Delta m_{\text{full}};\mu_{2},\sigma_{2}) \\ 
 &+(1-f_{1}-f_{2})f_{\text{student}}(\Delta m_{\text{full}};\Delta m_{0},\nu_{l},\nu_{h},\sigma_{\rm ave},\delta \sigma),
\end{align} where  $G(\Delta m_{\text{full}})$ is a Gaussian function defined in Appendix~A, and $f_{\text{student}}(\Delta m_{\text{full}})$ is obtained from the Student's t-distribution
\begin{equation}
f(t)=\frac{\Gamma(\nu/2+1/2)}{\Gamma(\nu/2)\sqrt{\nu\pi}}\cdot\left(1+\frac{t^{2}}{\nu}\right)^{(-\nu/2-1/2)},
\end{equation} where $\Gamma$ is the Gamma function. We define $t=(\Delta m_{\text{full}} - \Delta m_{0})/\sigma$ with $\Delta m_{0}$ and $\sigma$ the mean and width. In order to obtain the asymmetric t-function, the width parameter $\sigma$ and number of degrees of freedom $\nu$ are allowed to be different for the high and low sides of $\Delta m_{\text{full}}$. Widths for high and low sides of $\Delta m_{\text{full}}$ are then defined as:
$\sigma_{h}=\sigma_{\rm ave}+\delta \sigma$, and $\sigma_{l}=\sigma_{\rm ave}-\delta \sigma$,
and $\nu$ parameters for high and low sides are denoted as $\nu_h$ and $\nu_l$, respectively.  The bifurcated Student's t-function can then be defined as:
\begin{equation}
f_{\text{student}}(\Delta m_{\text{full}})=
\begin{cases}
\frac{r_{h}p_{h}}{\sqrt{\pi}} \cdot \left(1+\frac{(\frac{\Delta m_{\text{full}} - \Delta m_{0}}{\sigma_{h}})^{2}}{\nu_{h}}\right)^{(-\nu_{h}/2-1/2)} & \text{if $\Delta m_{\text{full}} - \Delta m_{0} \geq 0$,}
\\
\frac{r_{l}p_{l}}{\sqrt{\pi}} \cdot \left(1+\frac{(\frac{\Delta m_{\text{full}} - \Delta m_{0}}{\sigma_{l}})^{2}}{\nu_{l}}\right)^{(-\nu_{l}/2-1/2)} & \text{if $\Delta m_{\text{full}} - \Delta m_{0}<0$.}
\end{cases}
\end{equation} Auxiliary terms are defined as: 
\begin{equation}
p_{h}=\frac{\Gamma(\nu_{h}/2+1/2)}{\Gamma(\nu_{h}/2)\sqrt{\nu_{h}}|\sigma_{h}|},~
p_{l}=\frac{\Gamma(\nu_{l}/2+1/2)}{\Gamma(\nu_{l}/2)\sqrt{\nu_{l}}|\sigma_{l}|},~
r_{h}=\frac{2p_{l}}{p_{h}+p_{l}},~
r_{l}=\frac{2p_{h}}{p_{h}+p_{l}}~.
\end{equation} In total there are 11 shape parameters in the signal PDF, all of them are allowed to vary in the fit.
The background PDF is extracted from WS events, and is defined as:
\begin{equation}
f_{\rm bkg}(\Delta m_{\text{full}})=(1-f_3) \cdot f^{*}(\Delta m_{\text{full}})+f_3 \cdot BG(\Delta m_{\text{full}}),
\end{equation} where $f^{*}(\Delta m_{\text{full}})$ is defined in Eq.~\ref{eq:fstar}. We add a correction function, a bifurcated Gaussian, in order to have a better fit; the shape and the fraction of the bifurcated Gaussian is determined empirically from WS events. (We also use a background shape without this correction term to estimate the systematic uncertainty on the efficiency ratio.)
We include a ``signal" term in the fit to WS events to account for doubly-Cabibbo suppressed decays. 
\clearpage

\ifx\mcitethebibliography\mciteundefinedmacro
\PackageError{LHCb.bst}{mciteplus.sty has not been loaded}
{This bibstyle requires the use of the mciteplus package.}\fi
\providecommand{\href}[2]{#2}

\end{document}